# Vector representations and unit vector representations of fields — problems of understanding and possible teaching strategies

Christoph Hoyer, Raimund Girwidz
LMU München, Lehrstuhl für Didaktik der Physik

## Abstract

Vector fields are a highly abstract physical concept that is often taught using visualizations. Although vector representations are particularly suitable for visualizing quantitative data, they are often confusing, especially when describing real fields such as magnetic and electric fields, as the vector arrows can overlap. The present study examines vector understanding at the end of secondary education. In particular, the extent to which the geometry of the field can be derived from conventional unit vector representations and representations with centered unit vectors is examined. To support this understanding, two exercises were compared. The unirepresentational exercise argued within the conventional unit vector representation, while the multirepresentational exercise attempted to support the link between centered and conventional unit vectors.

The results show that almost all test subjects solved the items for generating vector representations correctly, but significant difficulties were encountered in interpreting vector representations. Drawing and interpreting vector representations therefore appear to be different skills that should be practiced intensively and in an integrated way. Furthermore, the learners recognized the field's geometry much more readily from centered unit vectors than from conventional unit vectors. Errors occur especially when interpreting the geometry of conventional unit vector representations of rotational fields and fields containing both sources and sinks, while the geometries of fields containing only sinks were interpreted quite well.

The comparison between the two training exercises showed that a promising approach to deepen students' understanding would be to use an exercise that contrasts the two representations and explains how to translate from one representation to the other, rather than describing the main elements of only a single representation. Finally, based on the results of the study, we propose a strategy for teaching vector representations in schools.

## I. Introduction

When dealing with invisible physical concepts and phenomena such as magnetic or electric fields, visualizations are essential to illustrate, communicate, and discuss their inherent properties. In education, qualitative methods such as field line drawings are typically used to characterize fields. While magnetic fields can be observed experimentally with iron filings or magnetic needles, electric fields can be demonstrated using semolina and castor oil or potassium permanganate [1]. Quantitative methods are rarely used for various reasons, not least because of the absence of experimental settings in which precise quantitative data can be obtained and the long time required to generate and visualize numerous measurements of the field's values, which are necessary to delineate its structure.

In the present day, digital media can reduce the time required for dataset generation or visualization and provide a quantitative alternative. For example, a web-based laboratory that allows precise measurement, various visualizations, and detailed analysis of the field of a permanent magnet is described in [2]. Further simulations that offer similar visualizations of electromagnetic fields can be found online, for example at https://phet.colorado.edu/ and http://www.didaktikonline.physik.uni-muenchen.de/programme/e_feld/E_Feld_min_en.html.

However, an open question remains regarding how to appropriately visualize quantitative data for students. Field line plots, which are commonly used for field visualizations, are unsuitable for visualizing accurate readings of measured data. Ordinary vector representations, in which the absolute value of the measured variable is represented by the vector's length, quickly become confusing, as the field's absolute values increase rapidly at positions closer to the charge or magnetic pole and the arrows overlap accordingly (see FIG. 1.).

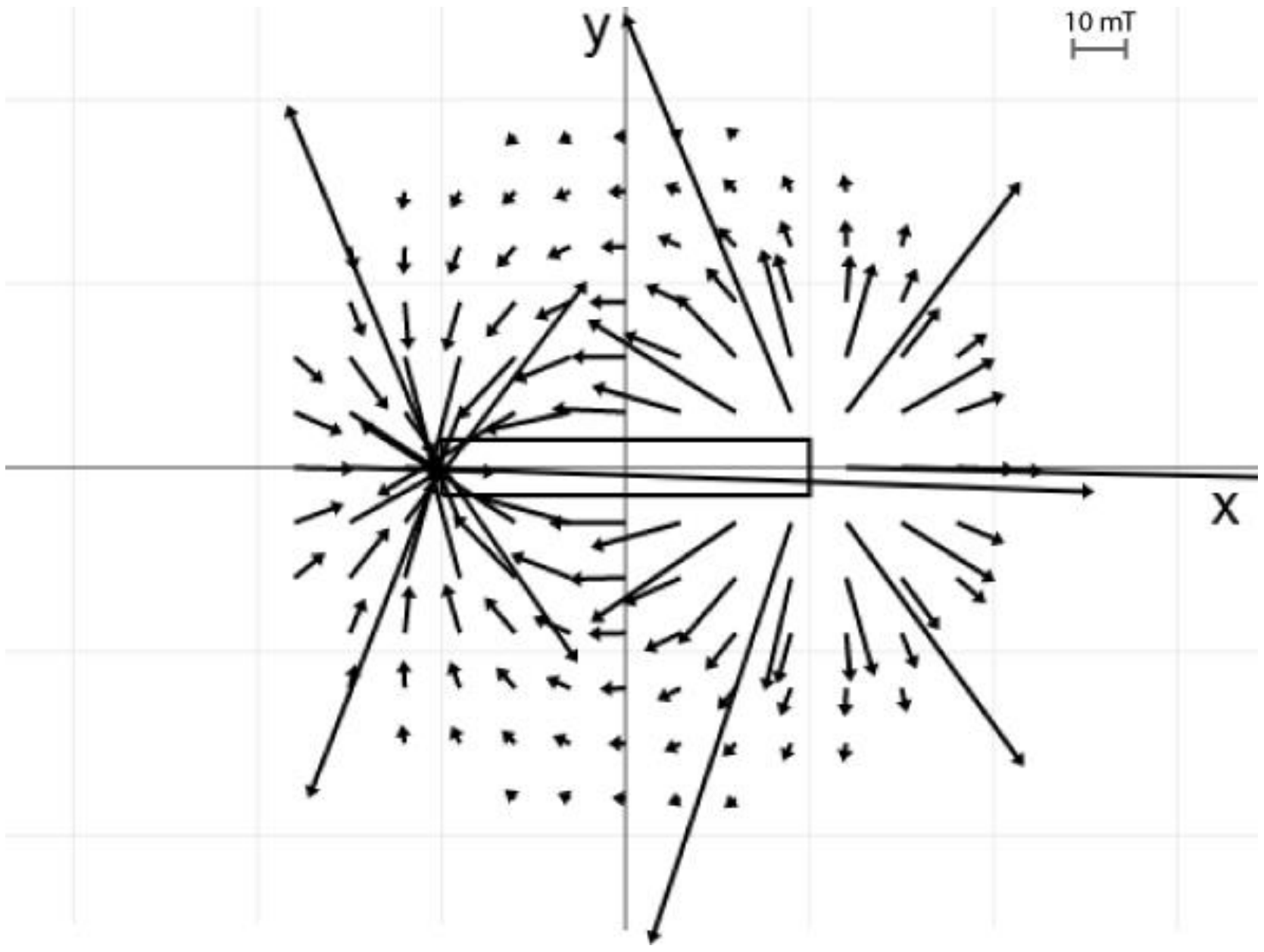

FIG. 1. Vector representation of the field of a permanent magnet. Due to the rapid increase in field strength closer to the poles, the vectors partially overlap, which makes this visualization confusing.

### A. Considering research results from cognitive science

According to Cognitive Load Theory [3-6], it is desirable to reduce the cognitive load related to the complexity of a representation to ensure sufficient cognitive resources are free to process the perceived information. Mayer & Moreno [7] proposed nine different approaches to reduce the cognitive load in multimedia learning environments—of these, segmenting information, pretraining, weeding of information, and signaling techniques are potentially suitable to reduce the cognitive load when learning from representations of vector fields. For example, the presentation can be segmented by separating information concerning the directions and magnitudes of the measured vectors into separate representations. This approach could be particularly useful when emphasizing single characteristics of the field during a pretraining phase. Furthermore, if only the directions or magnitudes of the measured values are relevant, by weeding out the other information the information load to be processed can be reduced. To subsequently link the representations, e.g., in combination with cross-representational questions, signaling techniques can be used.

A theoretical background for learning with more than one representation is described by the theory of learning with multiple representations [8]. Ainsworth states that learning with multiple external representations can be advantageous as they can fulfill different functions when combined [9]:

- Multiple representations can complement each other, both in terms of the content shown and the processes required for information processing.
- One representation can be used to constrain the scope for interpretation of another.
- Integrating information from different representations can facilitate a deeper understanding.

Separating the magnitude and direction of measured vector quantities into different representations therefore corresponds to the content branch of Ainsworth's proposed complementary function [8]. Overall, separating information into multiple representations is intended to reduce the information density of each representation and to stimulate deeper understanding.

It is highly challenging for teachers to select suitable visualizations of measured field values to ensure the best outcomes for learners. These representations should build on the learners' previous experiences but also present the field's characteristic features such that they can be correctly interpreted.

To address this issue, the present study aims to provide concrete insights into how the directionality of measured values of magnetic and electric fields can be visualized to ensure that the representations can be correctly understood by learners. Furthermore, promising potential approaches are identified and recommendations are made to help improve the understanding of directional representations of vectors.

## B. Background and related research on vector representations

The following section first provides an overview of how the concept of vectors is developed in traditional physics lectures. Problems with vector representations of physical fields are then discussed in more detail. Finally, the current research background and research gap are outlined.

Students first encounter vector quantities when they learn about physical forces in class. Forces have a point of action, a magnitude, and a direction, which are visualized by the starting point, length, and direction of the vector arrow, respectively. Later, this convention is used to visualize electric and magnetic fields. Here, the vectors are initially force arrows that visualize the force acting on a test charge in an electric field or the force acting on a test current in a magnetic field. Subsequently, based on these forces, the electric field strength and magnetic flux density are described as properties of defined points in space. Based on force vectors in mechanics, arrows are used to visualize these new vector quantities. The starting point of these field vectors indicates the position to which the value refers, while the magnitude and direction of the field are represented by the length and direction of the arrow, respectively.

In high school mechanics education, problems can usually be simplified so that the acting force is visualized by a single force arrow, e.g., the gravitational force is assumed to act at the object's center of gravity. If electromagnetic fields are visualized, however, many vector arrows are necessary, each starting from its reference point.

This type of representation can cause various comprehension difficulties during the learning process:

1) Vector arrows partially overlap, making the visualization confusing, e.g., when visualizing the magnetic flux density of a permanent magnet (see FIG. 1). Changes in the scaling of the length (e.g., logarithmic scaling) only help to a limited extent, especially as this approach can result in further misinterpretation.
2) Vector representations of fields can become problematic when interpreting the symmetry of the fields. Since the arrows start at the reference point and are plotted from there, asymmetries arise that can directly affect the interpretation of the field's presented geometry (see FIG. 2).

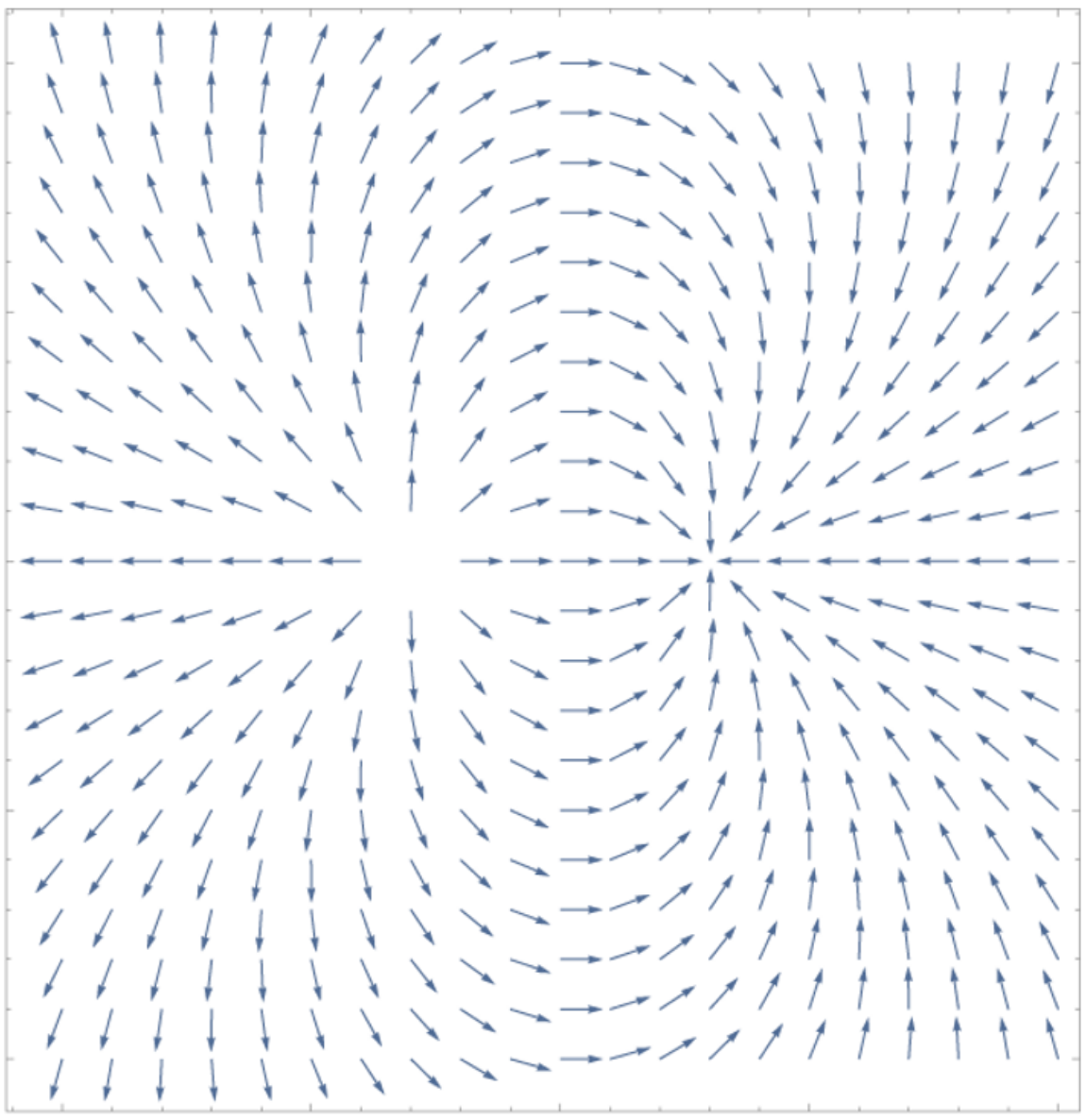

FIG. 2. Unit vector representation of the electric field surrounding two point charges with the same magnitudes and different signs. Due to the convention of how the vectors are drawn, asymmetries arise in the representation of the field geometries that could be misinterpreted.

As described in section I.A, overlap of vector arrows can be avoided by separating the information describing the vectors' magnitudes and directions into two different representations. One visualization shows the directions of the field, allowing its geometric structure to be interpreted, while another provides information about the field's magnitude.

In terms of deriving the geometry of the field from this type of directional representation, the question remains whether a conventional vector representation is suitable. One potential approach to better visualize the symmetries of the field would be to center the vectors at their reference points. However, to the best of our knowledge, no studies to date have provided any insights into whether this adapted form of representation provides any advantages in terms of understanding the field's geometry.

Previous studies have highlighted that two-dimensional field line images, e.g., those of point charges, which are often used in schools, are unsuitable for representing a vector field's strength as these images suffer from projection effects including equatorial clumping, false monopole moment, and boundary clumping [10]. To avoid these issues, Wolf et al. [10] suggested placing field vectors at the vertices of a square grid as an alternative form of representation; however, student problems with such vector representations have been previously reported. Based on the results of a developed "Vector Knowledge Test" [11], Knight concluded that students require explicit instruction and practice with the use of vectors. These results also led to further studies of physics students' understanding of vector addition, magnitude, and direction (e.g. [12,13]).

Problems with vector representations included that the vector quantity assigned to a point within a field is represented by an expanded spatial arrow that extends over a certain spatial area or the confusion between field lines and vector representations [13].

Gire and Price [14] highlight that vector representations contain an ambiguity in the meaning of space, which can cause difficulties for learners. Different points in the vector plot correspond to different locations in real space, and the distances between points in the vector plot indicate distances in real space; however, the distance between the starting point and tip of a vector is used as a measure of its absolute value. They also describe the misconception that a vector arrow is not only related to its reference point but to all points that it overlaps.

Based on previous studies [11,12, 15-30], Barniol and Zavala [31] developed a taxonomy describing the most frequent errors university students make after learning about vector concepts. Building on this, they developed a context-free multiple-choice vector concept test named the “test of understanding of vectors” (TUV). Four categories of the most frequently occurring incorrect answers could be identified, including difficulties in understanding the graphical properties of the direction, magnitude, and components of a vector [31]. More recent works presented a “Vector Field Representations” test (VFR) [32] and a concept test intended as a context-independent assessment of specific representational competence levels when working with field line images and vector field representations as well as when translating between these two representation types (RCFI) [33].

Bollen et al. [32] investigated student difficulties in terms of interpreting, constructing, and switching between vector field representations based on a target group of university students. The investigation showed that many misconceptions regarding vector representations remain even at the university level. One common misconception was identified during the construction of vector plots from their mathematical description, in which some learners incorrectly drew arrows centered on their reference point. This behavior was shown by students from a certain university and the cause of this phenomenon was not clear. Another misconception identified by Bollen et al. [32] was that the density of vector arrows was thought to describe the magnitude of the field. A possible explanation for this could be the activation of a cognitive structure described by [34], namely an intuitive knowledge element that Elby [34] calls WYSIWYG (“what-you-see-is-what-you-get”). For example, in the case of magnetic fields, this would mean that learners transfer previous experiences of visualizations using iron filings, in particular the increased accumulation of the filings at a magnet’s poles, to the vector representation of fields: this then leads to the misconception that closely adjacent vectors imply a stronger field. However, previous studies to date have not investigated whether there are differences in the perception of the transverse density and the longitudinal density of vectors.

In contrast to the VFR, the RCFI test addresses middle and high school students as well as entry-level university students. In the RCFI, no drawing tasks are required—although these are very good at identifying students' conceptions and competencies, they also make high generative demands on the learners [35]. For details, see [36] which describes a framework based on multimedia theory [37] that can be applied to the processes underlying drawing construction. Furthermore, the symbolic level of the vector field representation was omitted in the RCFI, since this aspect is not dealt with in-depth in schools. Overall, the results of the RCFI test showed that the field line concept, as understood by university students, is fraught with misconceptions [33]. It was proven that problems of understanding remain at the university level, the elimination of which was missed in physics classes at school.

The interpretation of unit vectors has not been extensively investigated in the past. In 2014, Barniol and Zavala [31] provided an overview of research into students' errors when dealing with different vector concepts. Research results on unit vectors in the Cartesian plane could not be found here. Their developed TUV contains an item in which the students are asked to find the unit vector in the direction of a given vector, and more than half the students could not solve this item correctly. The tasks in VFR that require unit vectors concern the construction of a possible mathematical expression for a given vector field presented graphically [32]. The graphical interpretation of unit vectors, on the other hand, is not required for this exercise.

Previous studies have already investigated various strategies to improve students' understanding of vector representations. For example, Klein et al. [38] conducted an eye-tracking study to compare the effectiveness of visual strategies to qualitatively interpret the divergence of graphical vector field representations. Their study found that learners benefited most when they were free to choose which of the two strategies to use. In a second study, Klein et al. [39] showed that the cognitive linking between the representations of vector fields and mathematical equations defining the vector field’s divergence and curl can be promoted by embedding visual cues in the instructional material. However, in both studies, the magnitudes at different locations in the vector fields only varied very slightly; hence, the representations were clear and unambiguous as the vectors did not overlap.

In contrast, in the case of real electromagnetic fields (especially near the poles), it is not usually possible to visualize the vector field without overlapping arrows. Unit vector representations are less problematic in this respect because overlap can be avoided due to the uniform vector length and corresponding spacing of the grid points. However, there has been no research to date into the interpretation and teaching of this type of directional representation. It is conceivable that the geometry of the fields may be better understood by learners using centered unit vectors rather than conventional vectors, but there are not yet sufficient empirical studies on this topic.

Additionally, there is no empirical evidence to date regarding what kind of exercise can improve understanding of conventional unit vector representations. Two promising approaches emerge from theory: first, based on the functions of multiple external representations [9], an exercise contrasting centered unit vectors and conventional unit vectors might help constrain the interpretation of the conventional vectors. Accordingly, such an exercise should aim to create global coherence [40] and an integrated cross-representational understanding. Second, in terms of the Single Concept Principle [41], an exercise could highlight core aspects of the unit vector representation and thus improve the understanding of this single representation in the sense of local coherence [40]. Since the learners would only have to concentrate on one representation type, according to the Cognitive Load Theory [3-6], resources in the working memory would be freed up for processing the essential information.

The present study aims to address these described research gaps. The findings of this work can help to anticipate problems in understanding vector field representations, which would allow appropriate measures to be prepared to correct these misunderstandings.

### C. Research questions

1. Are learners better able to derive the correct field line image from a conventional unit vector representation or a representation with centered unit vectors, based on the knowledge of vector fields they acquire during their school career?

2. What type of exercise is better suited to improve the interpretation of unit vector representations: a cross-representational exercise that relates conventional unit vectors to centered unit vectors or an exercise that uses a single representation?
3. How well can learners generate and interpret vector representations by the end of the 11th grade, and how do the two exercises influence comprehension?
4. Can learners identify areas where the magnitude of the field strength reaches its maximum based on unit vector representations of already-known field configurations, and, in addition, how do the two exercises influence this ability?

# II. Methods

## 1. Participants

A total of 124 students from eight different school classes participated in the study. The participants all attended the 11th grade of a Bavarian high school. The survey took place at the end of the school year. In Bavaria, representations of fields are discussed in schools for the last time in this grade; thus, at the time of the study, the learners had completed the full syllabus of field representations taught in schools.

## 2. Design

This study had two goals. First, vector understanding, including the interpretation of conventional unit vectors and centered unit vectors, was examined using a questionnaire. Second, a pre-post design was used to evaluate the effectiveness of two exercises, which both aimed to improve the generation and interpretation of vector representations. An intervention was conducted between the pre- and post-test. The two different intervention groups are called "multirepresentational" and "unirepresentational" (for details, see section II.3.a). The subjects were randomly assigned to the two groups. Figure 3 shows the study's design.

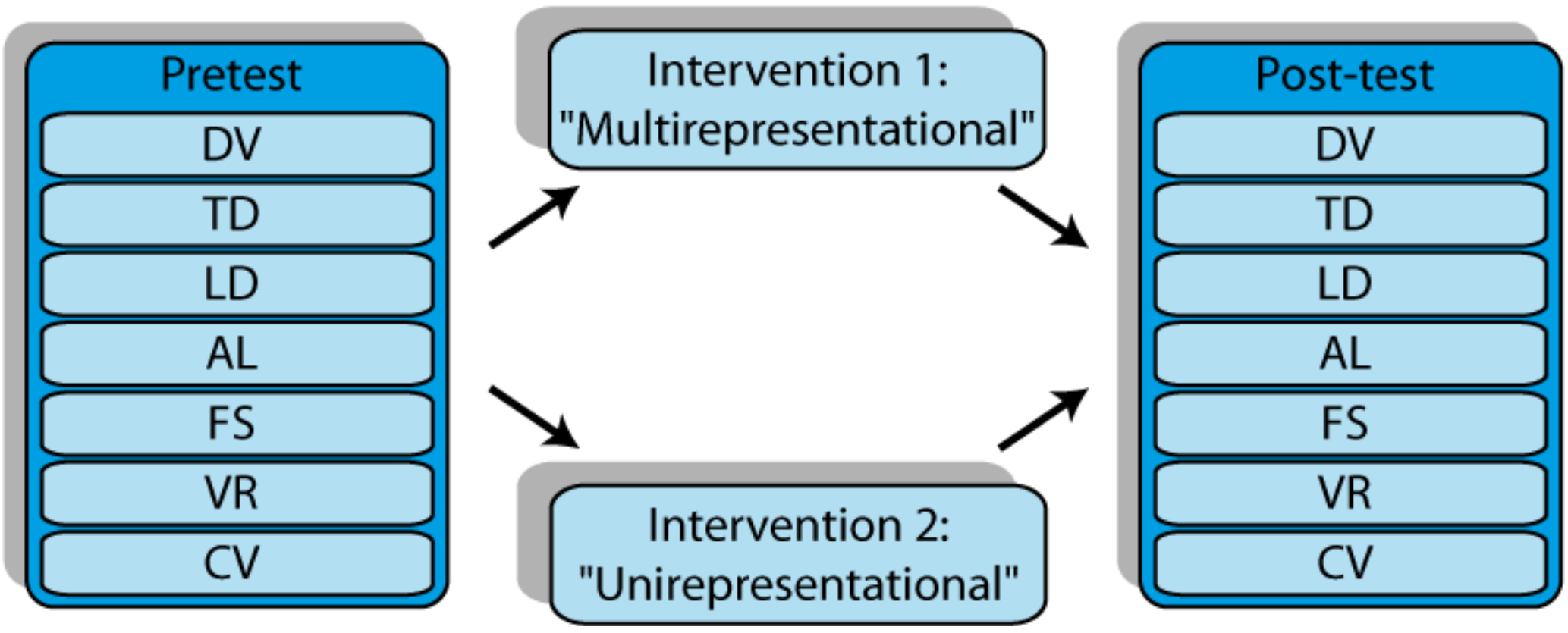

FIG. 3. Illustration of the study's design. The abbreviations of the individual test components are explained in detail in section II.3.b.

## 3. Materials

### *(a) Intervention*

Following the pretest, the intervention took place. Each participant worked on a learning environment about vector quantities that was presented on a tablet computer. The intervention took 20 minutes.

In both conditions, the learning environment comprised three parts. In the first part, participants reviewed aspects of vector quantities and their representations that they should have previously learned in school. Additionally, the relationship between conventional unit vectors, centered unit vectors, and field lines was discussed. The second part was the only one that differed in the two intervention groups. Here, a different strategy for interpreting conventional unit vectors was introduced in each intervention group (see below for a detailed description of the two strategies). In the third part of the application, the taught strategy had to be applied in two paper-and-pencil tasks to deepen understanding. In both tasks, the taught strategy first had to be applied to a conventional unit vector field representation, second, field lines had to be drawn into the representation, and third, the border of the measuring area had to be marked. After completing the task, the correct solution was presented on the tablet computer.

The following two subsections describe the strategies used to teach the derivation of field line images from conventional unit vector representations.

*Multirepresentational intervention*

Previous studies have reported that learning with multiple representations can lead to a deeper understanding of the subject matter [9]. Accordingly, an exercise that relates conventional and centered unit vectors is a potentially promising approach to promote improved understanding. Integrating both representations may also trigger coherence-building processes that lead to global coherence (see section I.B).

To promote the understanding of the conventional and centered unit vector representations, the strategy in this intervention aimed to establish a connection between the two representation types. As this approach uses multiple external representations to improve understanding, this condition is referred to as "multirepresentational" hereafter. In this intervention, an animation shows the conversion of a conventional unit vector into its centered representation; specifically, a conventional unit vector is extended backward by its length, resulting in a centered vector representation. The application argues that when drawing field lines in a conventional unit vector representation, the vectors in the studied area should first be converted to centered vectors.

*Unirepresentational intervention*

Working with two abstract representations at the same time can adversely affect the processing of the conveyed information due to cognitive overload (see I. A) [3-6]. By showing only one representation, the cognitive load could be reduced so that additional cognitive resources are available for processing the explanations associated with that representation. This approach is consistent with the Single Concept Principle (see I.B.). Therefore, the second intervention argues within the conventional vector representation. As the arguments are made only within one representation, this condition is referred to as "unirepresentational" hereafter. In the associated exercise, it was highlighted that when drawing field lines in a conventional unit vector representation, particular attention should be paid to the direction of the field on the axis.

*(b) Pre- and post-test*

The pretest and post-test consisted of the same items. In total these items addressed seven different constructs:

i) Drawing vectors (DV);
ii) interpreting the transverse density of vector arrows (TD);

iii) interpreting the longitudinal density of vector arrows (LD);
iv) interpreting the length of vector arrows (AL);
v) deriving information about field strength maxima from an associated directional representation (FS);
vi) interpreting the directions of a conventional unit vector representation (VR);
vii) interpreting the direction of a unit vector representation in which the vectors are centered at the reference point (CV).

In the following section, these eight constructs are described in detail.

### *(c) Drawing vectors (DV)*

The three items in this category test the extent to which learners can visualize magnetic flux density and electric field strength values using vectors. Each of the three items provides information about the location, direction, and magnitude of a vector quantity. The task involves visualizing this vector quantity in a preprinted coordinate system on the test sheet. The participants can achieve a maximum of three points per task. One point is awarded for each of the following three aspects: correct starting point, correct length, and correct direction of the vector arrow. To test the internal consistency, Cronbach's alpha was calculated for the pretest and post-test, with high values obtained for both the pretest ($\alpha$=0.921) and post-test ($\alpha$=0.911).

### *(d) Interpreting the transverse density of vectors (TD)*

Using two items, it was tested whether the transverse density of vector representations affects the participants' interpretation of the field strength. One of the items is shown in Fig. 4. Two vector plots were shown that differed in the density of vectors perpendicular to their direction. Each vector was of the same length. The task was to compare the field strengths of the two representations, and there were four answer alternatives: 1) field strength in representation 1 is higher, 2) field strength in representation 2 is higher, 3) both representations show a field of the same field strength, and 4) none of the possible answers is correct. To test the consistency of this two-item scale, Spearman–Brown coefficients were calculated for both the pretest (0.698) and the post-test (0.710).

| Which vector representation shows a field of higher electric field strength? | |
|---|---|
| 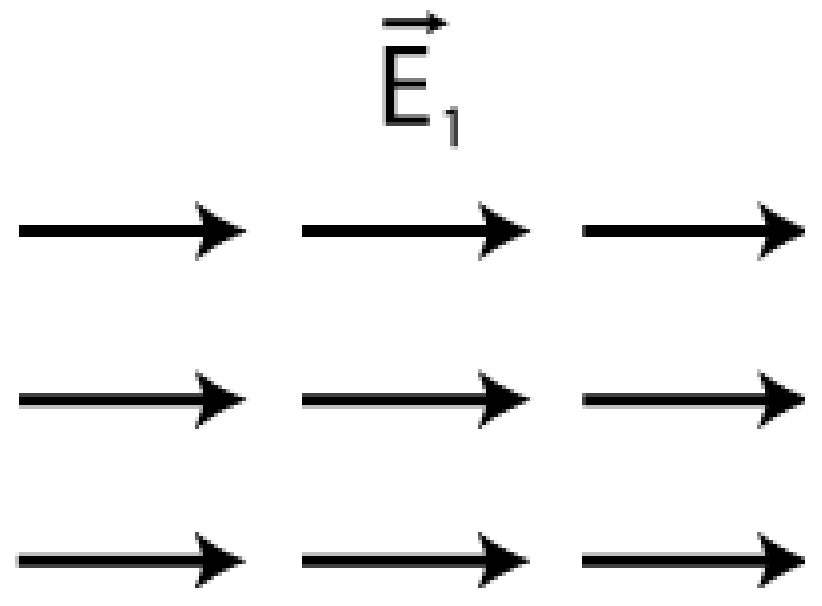 1. | 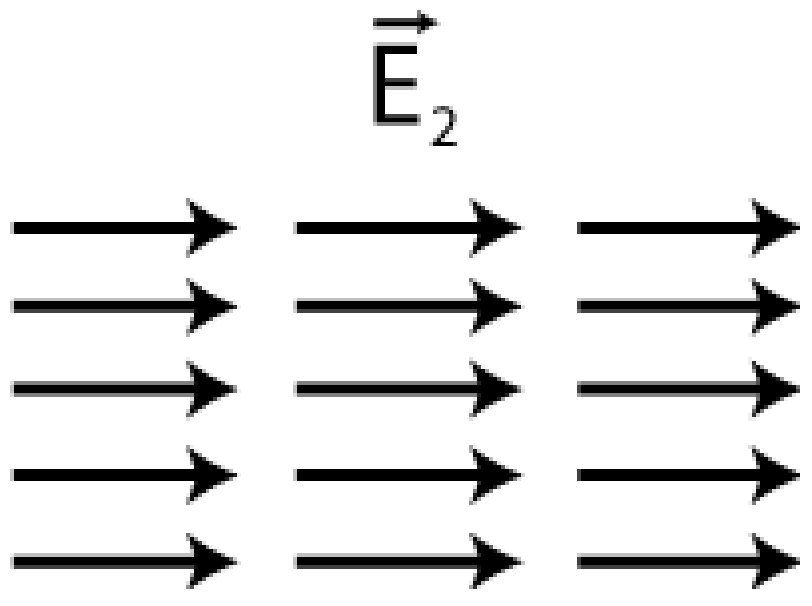 2. |
| ☐ Representation 1 | ☐ Representation 2 |
| ☐ The electric field strength in 1 and 2 is the same | ☐ None of the answer alternatives are correct |

FIG. 4. Example of an item regarding the interpretation of the transverse density of vectors.

### *(e) Interpreting the longitudinal density of vectors (LD)*

Two items were used to test if the longitudinal density of vectors influences the participants' interpretation of the field strength. One of these items is depicted in Fig. 5. Two vector plots were shown that differed in terms of the density at which the vectors were plotted in the direction of their vector arrows. All vectors were of equal length. The task was to compare the field strengths of the two representations, and there were four answer alternatives: 1) field strength in representation 1 is higher, 2) field strength in representation 2 is higher, 3) both representations show a field of the same field strength, and 4) none of the possible answers is correct. To test the consistency of this two-item scale, Spearman–Brown coefficients were again calculated for both the pretest (0.869) and the post-test (0.847).

| Which vector representation shows a field of higher electric field strength? | |
|---|---|
| 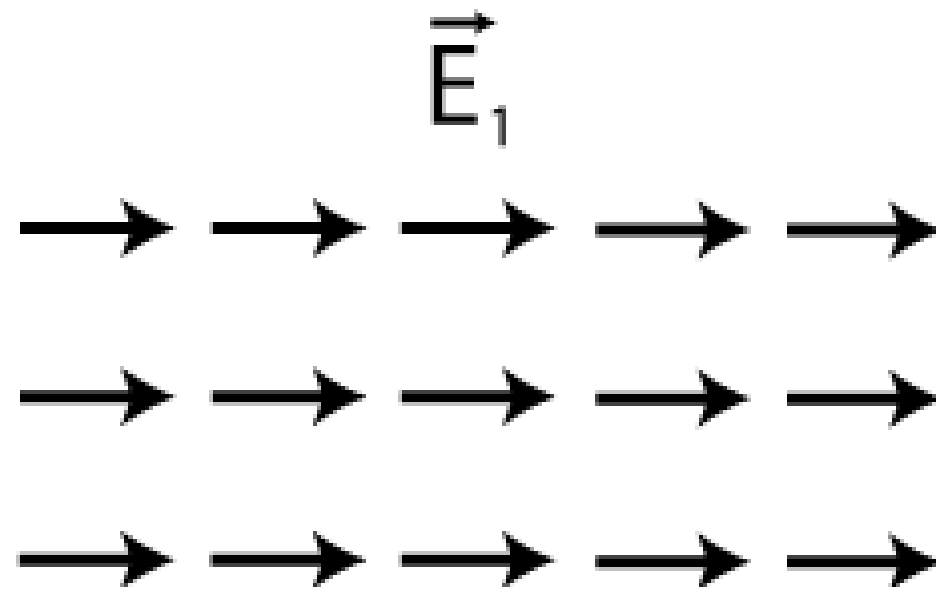  1. | 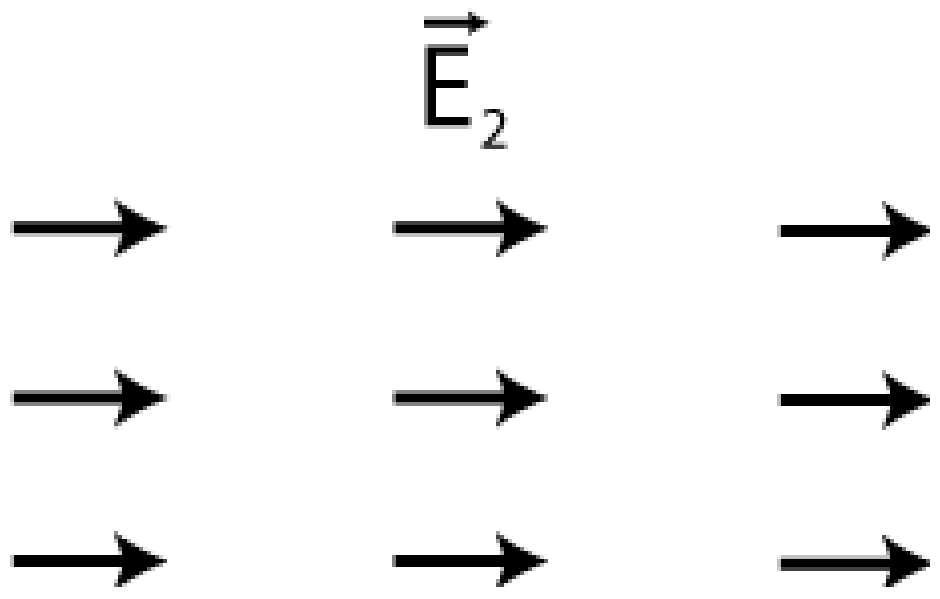  2. |
| ☐ Representation 1 | ☐ Representation 2 |
| ☐ The electric field strength in 1 and 2 is the same | ☐ None of the answer alternatives are correct |

FIG. 5. Example of an item regarding the interpretation of the longitudinal density of vectors.

### *(f) Interpretation of the length of vector arrows (AL)*

Another two-item scale was used to investigate whether the length of the displayed vectors is correctly interpreted as an indicator of field strength. Figure 6 shows one of the two items. Again, two vector plots were presented. Each consisted of the same number of vectors which all pointed in the same direction and were located at the same positions in the two diagrams. The only difference was that the arrows in the two representations differed in length. The participants were asked which of the two visualizations showed a higher field strength. They had to choose between four answer alternatives: 1) field strength in representation 1 is higher, 2) field strength in representation 2 is higher, 3) both representations show a field of the same field strength, and 4) none of the possible answers is correct. To test the consistency of this two-item scale Spearman–Brown coefficients were calculated for both the pretest (0.741) and the post-test (0.833).

| Which vector representation shows a field of higher electric field strength? | |
|---|---|
| 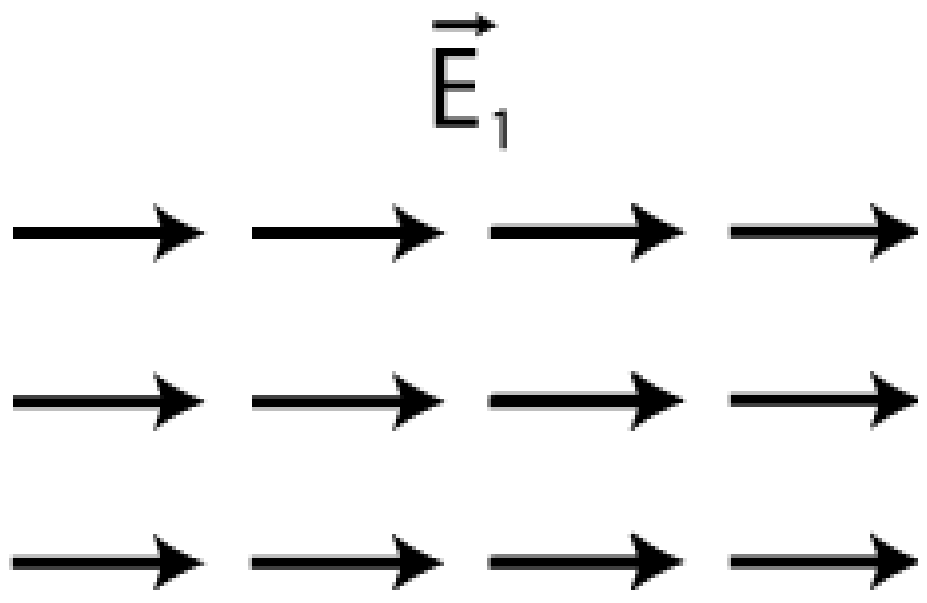  1. | 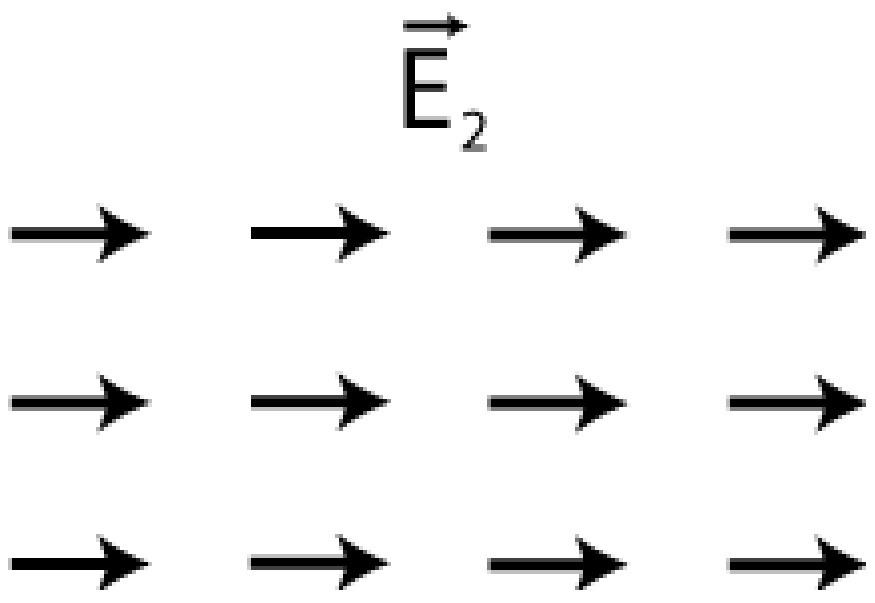  2. |
| ☐ Representation 1 | ☐ Representation 2 |
| ☐ The electric field strength in 1 and 2 is the same | ☐ None of the answer alternatives are correct |

FIG. 6. Example of an item regarding the interpretation of the length of vector arrows.

*(g) Deriving information about field strength maxima from a directional representation (FS)*

If working with known field configurations, based on previously acquired knowledge, the positions of field strength maxima can be deduced from the fields' directions. The two items in this scale test whether the participants can correctly compare field strength maxima in three given areas. One of the two items is depicted in Figure 7. The vector representations of the two items both show the field directions resulting from two point charges with the same absolute value but different signs. The two items differ only in terms of the position of the point charges. In the first item, the positive charge is on the left while the negative charge is on the right and vice versa in the second item. Two of the three marked areas in which the magnitudes of the field strengths are to be compared (areas 1 and 2 in Figure 7) contain a point charge. The third area (area 3 in Figure 7) does not contain a charge and is located between the two charges, outside the axis of symmetry. For each item, the participants can choose between four alternative answers: 1) the maximum of the absolute value of the field strength in area 1 is higher than in area 2, 2) the maximum of the absolute value of the field strength in area 2 is higher than in area 1, 3) the maximum of the absolute value of the field strength in area 1 is equal to that in area 2, while the maximum of the absolute value of the field strength in area 3 is smaller than both, and 4) the maxima of the absolute value of the field strength in areas 1, 2, and 3 are the same. Again, to test the consistency of this two-item scale, Spearman–Brown coefficients were calculated for both the pretest (0.931) and the post-test (0.950).

| |
|---|
| For the following directional field $E_{1max}$, $E_{2max}$, and $E_{3max}$ are the respective maxima of the absolute values of the electric field strength in areas 1), 2), and 3). Which statement is correct? |

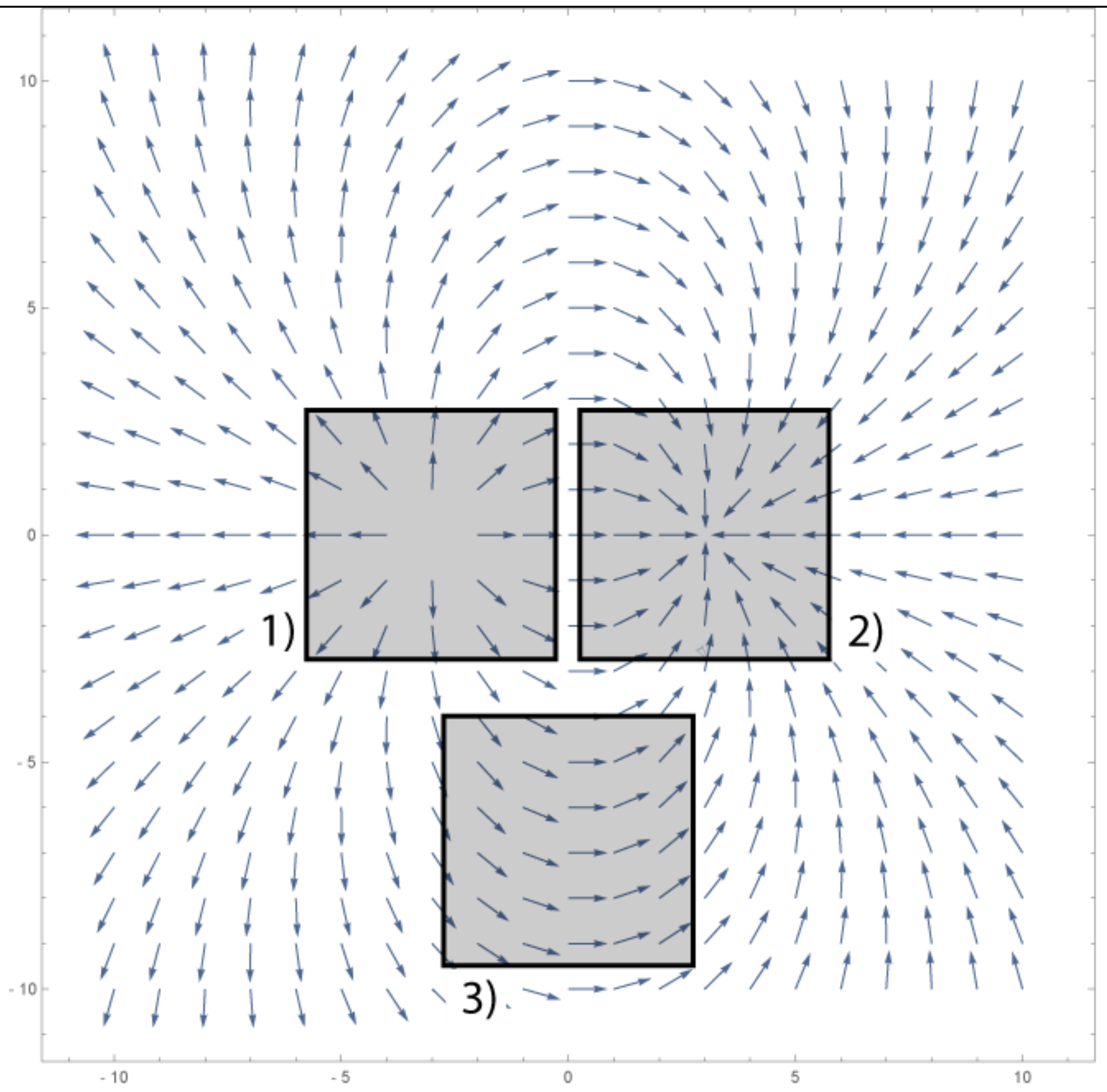

| | |
|---|---|
| ☐ $E_{1max} > E_{2max}$ | ☐ $E_{1max} < E_{2max}$ |
| ☐ $E_{1max} = E_{2max} > E_{3max}$ | ☐ $E_{1max} = E_{2max} = E_{3max}$ |

FIG. 7. Example of an item regarding interpreting the strength of a vector field from its directional representation.

### *(h) Interpreting the directions of a conventional unit vector representation (VR)*

10 items tested the participants' ability to relate the correct field line picture to a given unit vector representation. The 10 tested field configurations were:

i. Electric field of one point charge;
ii. electric field of two point charges of the same magnitude and same sign;
iii. electric field of two point charges of the same magnitude but different signs;
iv. electric field of two point charges of different magnitude and the same sign;
v. electric field of two point charges of different magnitude and different signs;
vi. magnetic field of a current-carrying wire;
vii. magnetic field of two current-carrying wires where the currents have the same magnitude and the same direction;
viii. magnetic field of two current-carrying wires where the currents have the same magnitude and opposing directions;
ix. magnetic field of two current-carrying wires where the currents have different magnitudes and the same direction;
x. magnetic field of two current-carrying wires where the currents have different magnitudes and opposing directions.

For each field configuration, the participants had to select the field line picture that best fits the displayed vector plot. In total, eight different answer alternatives were given. Distractors addressed different symmetries and properties of fields as well as different geometries of the borders of the measurement area. To test the internal consistency, Cronbach's alpha was determined for the pretest (0.560) and post-test (0.843). One of the 10 items is shown in Figure 8.

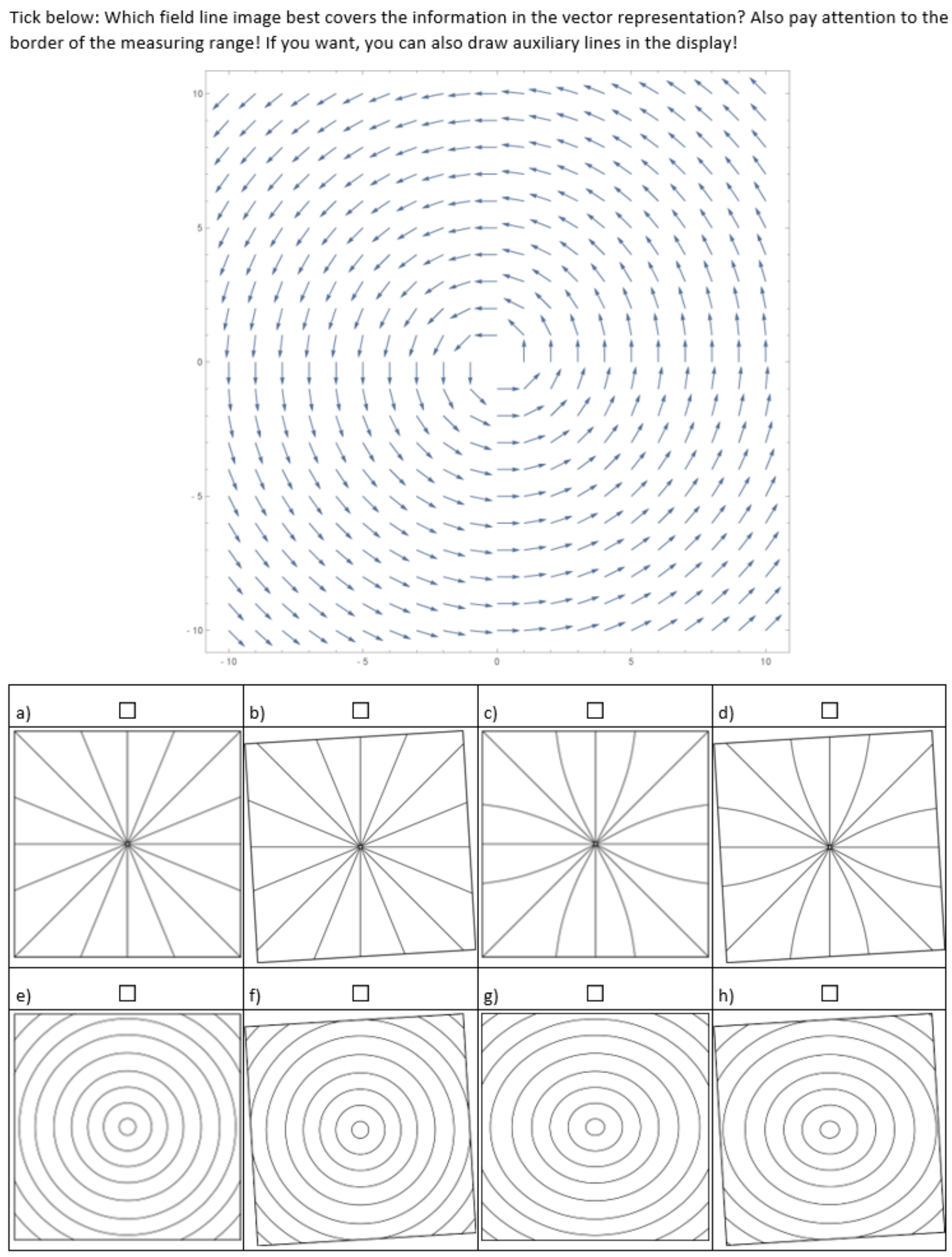

FIG. 8. Example of an item regarding the interpretation of the directions of a conventional unit vector representation.

### *(i) Interpretation of the directions of centered unit vectors (CV)*

A further 10 items tested the participants' ability to associate the correct field line picture with a given unit vector representation, where the vectors were centered at the point to which they belong. The 10 field configurations were the same as mentioned above (see II.3.h), and the same distractors were used. Again, participants had to choose from eight alternative field line pictures and select the one that best represents the given field. To test the internal consistency of the 10 items, Cronbach's alpha was calculated for the pretest (0.790) and post-test (0.714). One of the 10 items is shown in Figure 9.

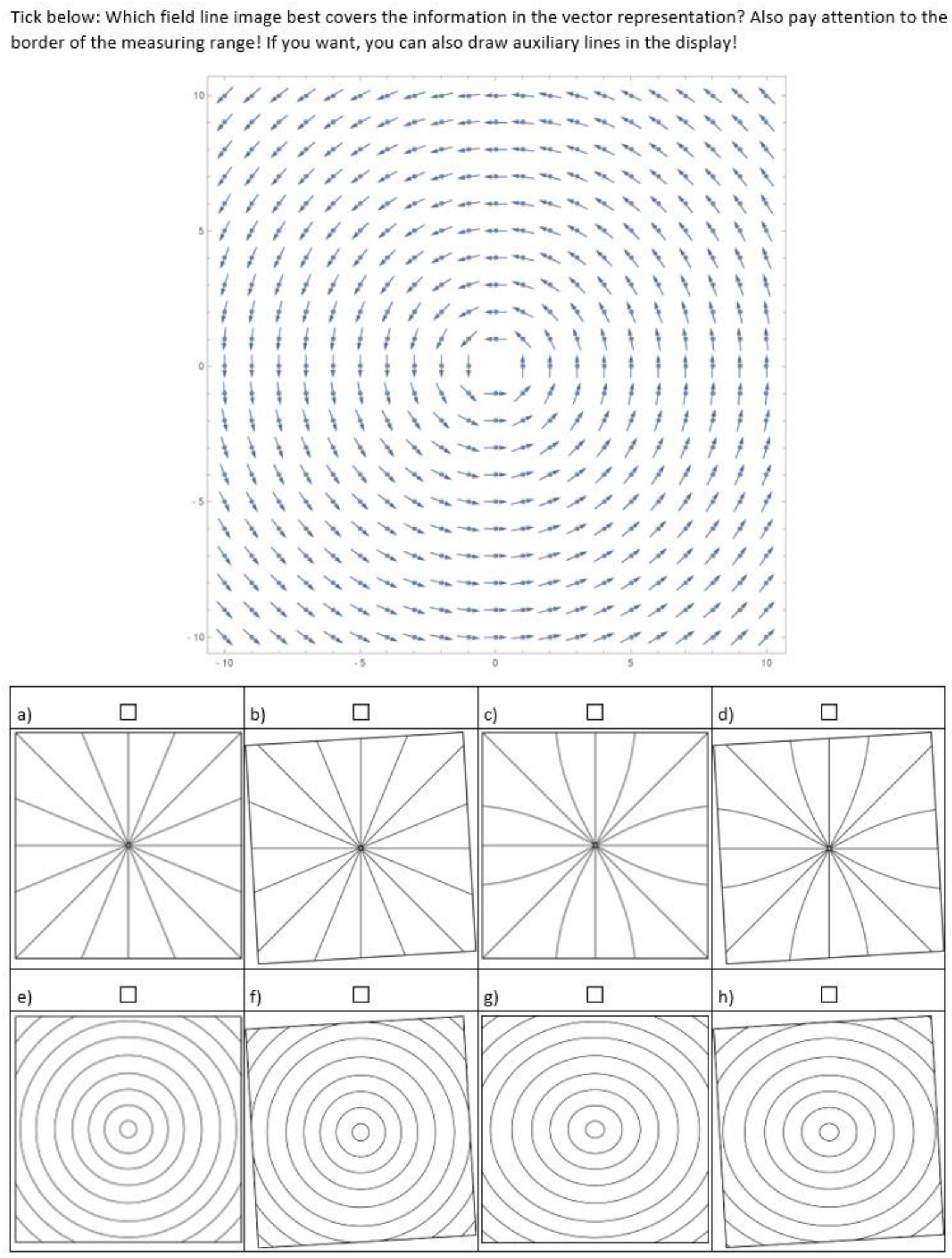

FIG. 9. Example of an item regarding the interpretation of the directions of a unit vector representation with centered vectors.

### 4. Procedure

A total of eight different 11th-grade classes participated in the study. In the 11th grade of the G8 curriculum in Bavaria, electrical and magnetic fields are taught in school for the last time. To ensure that all school-relevant content on magnetic and electric fields had already been studied, the examination was conducted at the end of the school year. First, the participants worked on the pretest for 30 minutes before being randomly assigned to one of the two intervention groups. During the intervention, participants worked through the learning environment on the tablet for 20 minutes. Finally, they had 30 minutes to complete the post-test.

### 5. Statistical Methods

#### *(a) Internal consistency of the test items*

In the pretest and post-test, different physical concepts regarding the graphical representations of vector fields are assessed (see section II.3.b). Some of these concepts are represented by three or more items, whereas others are represented by only two items for reasons of test efficiency. To test the internal consistency of the test components with three or more items,

Cronbach's alpha was calculated, while Spearman–Brown coefficients were used to assess the internal consistency of the two-item scales, as recommended by [42].

*(b) Exploratory factor analysis*

An exploratory factor analysis was performed to examine the structure of the questionnaire in more detail. To test whether the correlation matrix is significantly different from an identity matrix, Bartlett's test was conducted. Furthermore, the Kaiser–Meyer–Olkin measure of sampling adequacy was calculated to check how suitable the data are for an exploratory factor analysis. Because the data are not continuous, a mixed correlation matrix was calculated using the psych package in R. Since the factors are expected to correlate with each other, oblique rotation was performed using the oblimin rotation method. In addition, the Kaiser criterion and a scree plot were used to determine the number of factors.

*(c) Wilcoxon test and Mann–Whitney U-test*

Data were not normally distributed in either intervention group. Therefore, instead of computing paired sample t-tests, non-parametric Wilcoxon tests were calculated to examine the differences between the pre- and post-test for each group. Similarly, non-parametric Mann–Whitney U-tests were used to investigate differences between groups in terms of the improvement between the pre- and post-tests. In accordance with [43], the distributions of the dependent variable were checked for differences between the groups using Kolmogorov–Smirnov tests. To account for a multiple comparison problem, the results are corrected according to [44].

# III. Results

## 1. Exploratory factor analysis of the structure of the questionnaire

To test the questionnaire's structure, a factor analysis (minimum residual solution) was conducted on the 31 pretest items with oblique rotation. The Kaiser–Meyer–Olkin measure confirmed the adequacy of the data (KMO = 0.61; "mediocre", according to [45]). Bartlett's test of sphericity, $\chi^2$ (465) = 1782, $p < 0.001$, indicated that the correlations between the items were sufficiently large. An initial analysis of the eigenvalues of each component revealed that eight of the components had eigenvalues greater than the Kaiser criterion of 1.

The scree plot was ambiguous and suggested seven, eight, or eleven components. Since the content of the questionnaire was designed to capture seven constructs and this number is consistent with the scree plot (whereas Kaiser's criterion tends to overestimate the number of factors [46]), seven components were used for subsequent analysis.

Table 1 shows the factor loadings after rotation. The results reflect the originally intended structure relatively well:

- Factor 1 represents the ability to draw vectors.
- Factor 2 represents the understanding that a vector field's strength is not encoded in the transverse density of vector arrows.
- Factor 3 represents the understanding that a vector field's strength is not encoded in the longitudinal density of vector arrows.
- Factor 4 represents the interpretation of the length of vectors in terms of field strength.
- Factor 5 represents the ability to derive information about the absolute field strength value from a given unit vector representation.
- Factor 6 reflects the interpretation of unit vector representations.

- Factor 7 reflects the interpretation of centered unit vector representations.

However, factor 6 must be considered in more detail. The items that load particularly high on this factor all show rotational fields (CV1, CV5, CV7, and CV8; field configurations VI, IX, VII, and X, see section II.3.h). This factor thus appears to particularly reflect the ability to interpret conventional unit vector representations of rotational fields. Although not quite as clearly, items CV3, CV9, and CV10 also load on this factor. These items show electric fields with sinks and sources (CV3 and CV9; field configurations V and III, respectively) and a magnetic field caused by currents of the same magnitude and opposite direction (CV10; field configuration VIII). However, items CV2, CV4, and CV6 (field configurations I, II, and IV), which show field configurations only containing sinks, have almost no load on factor 6 but instead load on the factor regarding the interpretation of centered unit vectors (factor 7).

Table 2 presents the correlations between the identified factors. Interestingly, the factor relating to the ability to draw vectors (factor 1) shows almost no correlation with the three factors representing the ability to interpret vector representations (factors 2–4). In addition, the factors relating to the interpretation of the longitudinal and transverse density of the vectors are only weakly to moderately correlated.

| **Item** | Factor 1 | Factor 2 | Factor 3 | Factor 4 | Factor 5 | Factor 6 | Factor 7 |
|---|---|---|---|---|---|---|---|
| VZ1 | **1.10** | 0.00 | 0.03 | -0.05 | 0.00 | -0.08 | 0.05 |
| VZ2 | **1.11** | 0.06 | 0.03 | 0.06 | 0.00 | -0.03 | 0.01 |
| VZ3 | **0.67** | -0.07 | 0.05 | 0.04 | -0.14 | -0.18 | 0.10 |
| TD1 | 0.04 | **0.97** | 0.13 | 0.03 | -0.03 | 0.06 | 0.10 |
| LD1 | -0.10 | 0.34 | **0.78** | -0.03 | 0.26 | -0.13 | -0.05 |
| AL1 | 0.07 | **0.48** | 0.02 | **0.69** | -0.07 | -0.02 | -0.12 |
| TD2 | 0.01 | **0.65** | 0.21 | 0.32 | 0.02 | -0.08 | -0.09 |
| LD2 | 0.08 | 0.08 | **0.93** | 0.03 | 0.06 | 0.06 | 0.06 |
| AL2 | 0.04 | 0.02 | -0.02 | **0.83** | -0.03 | 0.16 | 0.12 |
| FS1 | 0.10 | 0.01 | 0.22 | -0.08 | **0.81** | -0.02 | 0.16 |
| FS2 | 0.12 | -0.11 | 0.30 | -0.04 | **0.78** | -0.19 | 0.08 |
| VR1 | 0.11 | -0.21 | 0.06 | 0.27 | 0.25 | **0.44** | 0.12 |
| VR2 | -0.01 | -0.04 | 0.21 | 0.33 | 0.15 | -0.03 | 0.24 |
| VR3 | **0.54** | 0.27 | -0.39 | -0.31 | **0.41** | 0.30 | 0.06 |
| VR4 | 0.05 | -0.11 | 0.20 | 0.16 | 0.25 | -0.08 | **0.55** |
| VR5 | 0.12 | 0.00 | 0.08 | -0.08 | -0.04 | **0.84** | 0.10 |
| VR6 | 0.26 | -0.29 | 0.20 | -0.20 | -0.21 | 0.08 | 0.39 |
| VR7 | -0.04 | -0.15 | 0.00 | 0.17 | 0.05 | **0.82** | -0.11 |
| VR8 | -0.17 | 0.11 | -0.19 | 0.12 | -0.10 | **0.94** | 0.04 |
| VR9 | **0.65** | -0.25 | -0.12 | 0.23 | 0.30 | 0.31 | **-0.41** |
| VR10 | -0.06 | 0.08 | -0.39 | 0.03 | **0.65** | 0.29 | -0.13 |
| CV1 | -0.04 | 0.24 | -0.01 | -0.12 | 0.18 | 0.13 | **0.71** |
| CV2 | 0.12 | 0.14 | 0.11 | 0.08 | -0.13 | -0.07 | **0.74** |
| CV3 | -0.02 | 0.02 | -0.11 | 0.02 | 0.20 | -0.15 | **0.69** |
| CV4 | -0.24 | -0.28 | 0.00 | 0.06 | 0.22 | -0.06 | **0.73** |
| CV5 | -0.17 | -0.13 | -0.02 | 0.19 | **0.42** | -0.16 | **0.49** |
| CV6 | 0.02 | 0.10 | -0.07 | 0.07 | 0.02 | -0.07 | **0.69** |
| CV7 | 0.17 | 0.04 | 0.00 | -0.06 | -0.12 | 0.04 | **0.86** |
| CV8 | -0.28 | 0.00 | -0.07 | -0.01 | 0.21 | -0.17 | **0.49** |
| CV9 | 0.04 | -0.10 | -0.31 | 0.27 | -0.03 | **-0.69** | 0.32 |
| CV10 | 0.12 | -0.21 | 0.38 | -0.02 | -0.19 | 0.00 | **0.57** |
| Eigenvalues | 4.05 | 2.37 | 2.65 | 1.89 | 2.74 | 3.76 | 5.34 |
| Proportion of Variance | 0.13 | 0.08 | 0.09 | 0.06 | 0.09 | 0.12 | 0.17 |

Table 1. Results of the exploratory factor analysis. Factor loadings over 0.4 are printed in bold.

| | Factor 1 | Factor 2 | Factor 3 | Factor 4 | Factor 5 | Factor 6 | Factor 7 |
|---|---|---|---|---|---|---|---|
| Factor 1 | 1.00 | 0.04 | 0.00 | -0.04 | 0.21 | 0.16 | -0.03 |
| Factor 2 | 0.04 | 1.00 | 0.16 | 0.07 | 0.00 | 0.08 | -0.04 |
| Factor 3 | 0.00 | 0.16 | 1.00 | 0.01 | 0.02 | -0.30 | 0.19 |
| Factor 4 | -0.04 | 0.07 | 0.01 | 1.00 | -0.03 | 0.02 | -0.02 |
| Factor 5 | 0.21 | 0.00 | 0.02 | -0.03 | 1.00 | 0.12 | 0.11 |
| Factor 6 | 0.16 | 0.08 | -0.30 | 0.02 | 0.12 | 1.00 | -0.33 |
| Factor 7 | -0.03 | -0.04 | 0.19 | -0.02 | 0.11 | -0.33 | 1.00 |

Table 2. Correlation matrix between the factors.

## 2. Conventional vs. centered unit vector plots

To answer the first research question, which relates to whether field geometries can be more easily recognized from conventional or centered unit vector representations, the results of the VR and CV items of the pretest were evaluated.

The participants achieved an average score of 3.15 out of 10 possible points when answering the items regarding conventional unit vectors (VR). In contrast, in the tasks with centered unit vectors (CV), they scored an average of 8.23 out of 10 points. The Wilcoxon signed-rank test results reveal significant differences in the median scores achieved in the two task categories ($Z = 9.138$, $p < 0.01$, $r = 0.82$, $R^2 \approx 0.67$). According to Cohen [47], this corresponds to a large effect size.

| Field Config. | I | II | III | IV | V | VI | VII | VIII | IX | X |
|---|---|---|---|---|---|---|---|---|---|---|
| Item Nr. | 2 | 4 | 9 | 6 | 3 | 1 | 7 | 10 | 5 | 8 |
| Correct | 0.73 | 0.81 | 0.06 | 0.78 | 0.06 | 0.21 | 0.13 | 0.12 | 0.16 | 0.09 |
| Wrong geometry | 0.20 | 0.12 | 0.05 | 0.17 | 0.06 | 0.11 | 0.03 | 0.00 | 0.04 | 0.03 |
| Wrong border | 0.05 | 0.06 | 0.54 | 0.04 | 0.55 | 0.52 | 0.77 | 0.74 | 0.59 | 0.77 |
| Wrong geometry & border | 0.02 | 0.00 | 0.35 | 0.01 | 0.33 | 0.15 | 0.06 | 0.14 | 0.19 | 0.10 |

Table 3. Detailed analysis of the pretest answers relating to the VR items (i.e., interpretation of conventional unit vector representations). The table shows the frequency of the correct answer and the frequency at which the geometry, border, or both were estimated incorrectly. The values are quoted as fractions.

| Field Config. | I | II | III | IV | V | VI | VII | VIII | IX | X |
|---|---|---|---|---|---|---|---|---|---|---|
| Item Nr. | 2 | 7 | 6 | 10 | 9 | 1 | 3 | 5 | 8 | 4 |
| Correct | 0.86 | 0.84 | 0.86 | 0.72 | 0.77 | 0.77 | 0.89 | 0.85 | 0.82 | 0.85 |
| Wrong geometry | 0.12 | 0.09 | 0.10 | 0.21 | 0.19 | 0.22 | 0.06 | 0.10 | 0.12 | 0.10 |
| Wrong border | 0.02 | 0.05 | 0.02 | 0.05 | 0.02 | 0.02 | 0.06 | 0.02 | 0.05 | 0.04 |
| Wrong geometry & border | 0.00 | 0.02 | 0.01 | 0.02 | 0.01 | 0.00 | 0.00 | 0.01 | 0.00 | 0.01 |

Table 4. Detailed analysis of the pretest answers relating to the CV items (i.e., interpretation of centered unit vector representations). The values are quoted as fractions.

For the items with conventional (VR) and centered unit vectors (CV), Tables 3 and 4, respectively, show the frequency of the correct answers and the frequency at which the geometry, border, or both were estimated incorrectly. As shown, the correct answer frequency was similar for field configurations I, II, and IV in both conventional and centered unit vector representations. The most common source of error in these three configurations was that a different field geometry was identified, but in most cases, the border area of the unit vector representation was correctly interpreted (see Table 3).

For all other field configurations, better results were achieved from the centered unit vector representations. The most common source of error in the conventional unit vector representations was that the border of the displayed unit vector representation was not correctly recognized. In addition, both the border and geometry were quite commonly misinterpreted. More rarely, mistakes were made regarding the geometry of the field while the shape of the representation's border was correctly understood (see Table 3).

The items with centered unit vectors were consistently answered very well. If there were errors, these were mostly because the field's geometry was not interpreted correctly. The learners very rarely misinterpreted the border area in the centered unit vector representations (see Table 4).

3. Multirepresentational vs. unirepresentational exercise to improve interpretation of unit vectors

To compare the learning effectiveness of the two interventions with regard to the correct interpretation of conventional unit vector representations and representations with centered unit vectors, within and between differences were examined. To investigate the impact of the intervention for each group, Wilcoxon tests examined the difference between pretest and post-test scores when interpreting conventional and centered unit vector representations. A Mann–Whitney U-test was used to test for group differences in the difference scores between pre- and post-test in the interpretation of conventional and centered unit vectors.

The difference scores of the items with conventional unit vectors showed the same distribution, Kolmogorov-Smirnov $p > 0.05$. For the items with centered unit vectors, the distributions of the difference scores differed between the two groups, Kolmogorov-Smirnov $p < 0.05$.

In both the "multirepresentational" ($z = 5.646$, $p < 0.05$) and "unirepresentational" groups ($z = 4.905$, $p < 0.05$), the interpretation of conventional unit vectors improved significantly from pre- to post-test. However, for the items concerning centered unit vectors, only the "multirepresentational" group ($z = 2.204$, $p < 0.05$) showed a significant improvement, whereas no significant improvement was identified in the "unirepresentational" group ($z = -1.248$, $p > 0.05$).

In terms of the Mann–Whitney U-test used to examine the difference scores between pre- and post-test when working with conventional unit vectors, there were no significant differences between the "multirepresentational" (*Mdn* = 3.0) and "unirepresentational" (*Mdn* = 2.5) groups ($U = 1756.5$, $Z = -0.832$, $p > 0.5$). However, when investigating the difference scores in the "multirepresentational" ($M_{rank} = 69.57$) and "unirepresentational" ($M_{rank} = 55.43$) groups when working with centered unit vectors, there are clear differences ($U = 1483.5$, $Z = -2.282$, $p < 0.5$, $r = -0.29$, $R^2 \approx 0.08$). It should be noted that there are some potential ceiling effects as the tasks concerning the interpretation of centered unit vectors were often answered correctly in the pretest (particularly in the "unirepresentational" group; see Tables 7–8). In summary, in the part of the intervention which aims to improve the interpretation of vector representations, the multirepresentational strategy that relates centered and conventional unit vector representations achieved superior outcomes.

In all tasks regarding the interpretation of conventional unit vectors, the results increased from the pretest to the post-test. An analysis of the errors made in individual items in the pre- and post-test (see Tables 5–6) reveals that after the intervention, fewer errors were made when interpreting the border of the conventional unit vector representation. However, mistakes were still frequently made, especially in field configurations III, V, VI, VII, VIII, IX, and X.

| Field Config. | I | II | III | IV | V | VI | VII | VIII | IX | X |
|---|---|---|---|---|---|---|---|---|---|---|
| Item Nr. | 2 | 4 | 9 | 6 | 3 | 1 | 7 | 10 | 5 | 8 |
| Pretest | | | | | | | | | | |
| Correct | 0.74 | 0.74 | 0.10 | 0.69 | 0.08 | 0.24 | 0.13 | 0.11 | 0.21 | 0.10 |
| Wrong geometry | 0.18 | 0.18 | 0.06 | 0.24 | 0.10 | 0.16 | 0.05 | 0.00 | 0.08 | 0.06 |
| Wrong border | 0.05 | 0.08 | 0.47 | 0.05 | 0.53 | 0.39 | 0.74 | 0.68 | 0.52 | 0.69 |
| Wrong geometry & border | 0.03 | 0.00 | 0.37 | 0.02 | 0.29 | 0.19 | 0.08 | 0.21 | 0.18 | 0.15 |
| Post-test | | | | | | | | | | |
| Correct | 0.81 | 0.89 | 0.37 | 0.89 | 0.47 | 0.53 | 0.58 | 0.60 | 0.52 | 0.56 |
| Wrong geometry | 0.16 | 0.10 | 0.29 | 0.11 | 0.16 | 0.10 | 0.02 | 0.08 | 0.08 | 0.08 |
| Wrong border | 0.00 | 0.00 | 0.26 | 0.00 | 0.19 | 0.31 | 0.37 | 0.29 | 0.24 | 0.32 |
| Wrong geometry & border | 0.02 | 0.02 | 0.08 | 0.00 | 0.18 | 0.06 | 0.03 | 0.03 | 0.16 | 0.03 |

Table 5. Pretest and post-test results of the single VR items in the "multirepresentational" group. The values are quoted as fractions.

| Field Config. | I | II | III | IV | V | VI | VII | VIII | IX | X |
|---|---|---|---|---|---|---|---|---|---|---|
| Item Nr. | 2 | 4 | 9 | 6 | 3 | 1 | 7 | 10 | 5 | 8 |
| Pretest | | | | | | | | | | |
| Correct | 0.73 | 0.89 | 0.02 | 0.87 | 0.03 | 0.18 | 0.13 | 0.13 | 0.11 | 0.08 |
| Wrong geometry | 0.23 | 0.06 | 0.03 | 0.10 | 0.03 | 0.06 | 0.02 | 0.00 | 0.00 | 0.00 |
| Wrong border | 0.05 | 0.05 | 0.61 | 0.03 | 0.56 | 0.65 | 0.81 | 0.81 | 0.66 | 0.85 |
| Wrong geometry & border | 0.00 | 0.00 | 0.34 | 0.00 | 0.37 | 0.11 | 0.05 | 0.06 | 0.21 | 0.06 |
| Post-test | | | | | | | | | | |
| Correct | 0.79 | 0.94 | 0.37 | 0.95 | 0.48 | 0.35 | 0.47 | 0.52 | 0.45 | 0.50 |
| Wrong geometry | 0.18 | 0.00 | 0.18 | 0.05 | 0.11 | 0.19 | 0.02 | 0.10 | 0.10 | 0.03 |
| Wrong border | 0.02 | 0.06 | 0.32 | 0.00 | 0.31 | 0.42 | 0.50 | 0.32 | 0.32 | 0.42 |
| Wrong geometry & border | 0.00 | 0.00 | 0.13 | 0.00 | 0.08 | 0.03 | 0.02 | 0.06 | 0.13 | 0.05 |

Table 6. Pretest and post-test results of the single VR items in the “unirepresentational” group. The values are quoted as fractions.

| Field Config. | I | II | III | IV | V | VI | VII | VIII | IX | X |
|---|---|---|---|---|---|---|---|---|---|---|
| Item Nr. | 2 | 7 | 6 | 10 | 9 | 1 | 3 | 5 | 8 | 4 |
| Pretest | | | | | | | | | | |
| Correct | 0.79 | 0.77 | 0.79 | 0.68 | 0.73 | 0.69 | 0.85 | 0.82 | 0.76 | 0.79 |
| Wrong geometry | 0.19 | 0.13 | 0.18 | 0.27 | 0.24 | 0.27 | 0.08 | 0.15 | 0.18 | 0.15 |
| Wrong border | 0.02 | 0.06 | 0.02 | 0.05 | 0.02 | 0.03 | 0.06 | 0.03 | 0.05 | 0.06 |
| Wrong geometry & border | 0.00 | 0.03 | 0.02 | 0.00 | 0.00 | 0.00 | 0.00 | 0.00 | 0.00 | 0.00 |
| Post-test | | | | | | | | | | |
| Correct | 0.95 | 0.94 | 0.79 | 0.77 | 0.69 | 0.79 | 0.92 | 0.95 | 0.87 | 0.92 |
| Wrong geometry | 0.05 | 0.06 | 0.18 | 0.19 | 0.26 | 0.21 | 0.05 | 0.02 | 0.10 | 0.08 |
| Wrong border | 0.00 | 0.00 | 0.03 | 0.03 | 0.02 | 0.00 | 0.03 | 0.02 | 0.02 | 0.00 |
| Wrong geometry & border | 0.00 | 0.00 | 0.00 | 0.00 | 0.03 | 0.00 | 0.00 | 0.02 | 0.02 | 0.00 |

Table 7. Pretest and post-test results of the single CV items in the “multirepresentational” group. The values are quoted as fractions.

| Field Config. | I | II | III | IV | V | VI | VII | VIII | IX | X |
|---|---|---|---|---|---|---|---|---|---|---|
| Item Nr. | 2 | 7 | 6 | 10 | 9 | 1 | 3 | 5 | 8 | 4 |
| Pretest | | | | | | | | | | |
| Correct | 0.94 | 0.90 | 0.94 | 0.76 | 0.81 | 0.84 | 0.92 | 0.89 | 0.89 | 0.90 |
| Wrong geometry | 0.05 | 0.05 | 0.03 | 0.15 | 0.15 | 0.16 | 0.03 | 0.06 | 0.06 | 0.06 |
| Wrong border | 0.02 | 0.03 | 0.02 | 0.05 | 0.03 | 0.00 | 0.05 | 0.02 | 0.05 | 0.02 |
| Wrong geometry & border | 0.00 | 0.00 | 0.00 | 0.05 | 0.02 | 0.00 | 0.00 | 0.02 | 0.00 | 0.02 |

| Post-test | | | | | | | | | | |
|---|---|---|---|---|---|---|---|---|---|---|
| Correct | 0.85 | 0.92 | 0.85 | 0.84 | 0.79 | 0.82 | 0.92 | 0.87 | 0.79 | 0.85 |
| Wrong geometry | 0.11 | 0.06 | 0.10 | 0.15 | 0.11 | 0.11 | 0.05 | 0.08 | 0.18 | 0.10 |
| Wrong border | 0.02 | 0.02 | 0.03 | 0.00 | 0.06 | 0.05 | 0.02 | 0.03 | 0.03 | 0.05 |
| Wrong geometry & border | 0.02 | 0.00 | 0.02 | 0.02 | 0.03 | 0.02 | 0.02 | 0.02 | 0.00 | 0.00 |

Table 8. Pretest and post-test results of the single CV items in the "unirepresentational" group. The values are quoted as fractions.

4. Multirepresentational vs. unirepresentational exercises to foster the abilities to generate and interpret vector representations

The statistical data for the items related to the generation and interpretation of vector representations are presented in Tables 9–17. In general, no significant differences in the improvement from pre- to post-test were identified between the two intervention groups for each of the item categories (Table 9). Both interventions were thus similarly efficient in improving the generation and interpretation of vector representations.

| | Multirepresentational | | Unirepresentational | | *U* | *Z* | *p* | *p_korr* | *r* |
|---|---|---|---|---|---|---|---|---|---|
| | Mean | SD | Mean | SD | | | | | |
| DV | 0.6290 (7.0 %) | 2.04245 (22.7 %) | 0.5000 (5.6 %) | 1.70582 (19.0 %) | 1843.0 | -0.498 | >0.05 | >0.05 | -0.04 |
| TD | 0.2097 (10.5 %) | 0.63082 (31.5 %) | 0.3710 (18.6 %) | 0.83438 (41.7 %) | 1773.5 | -0.928 | >0.05 | >0.05 | -0.08 |
| LD | 0.1613 (8.1 %) | 0.70580 (35.3 %) | 0.2258 (11.3 %) | 0.94815 (47.4 %) | 1884.0 | -0.237 | >0.05 | >.05 | -0.02 |
| AL | 0.1290 (6.5 %) | 1.15210 (57.6 %) | 0.4516 (22.6 %) | 1.12610 (56.3 %) | 1663.0 | -1.398 | >0.05 | >0.05 | -0.12 |

Table 9. Between-group comparison of the improvement in the items regarding the generation and interpretation of vector plots. For better interpretation, percentages are given in brackets.

A closer inspection of the pre- and post-test scores within the two groups (Tables 10–11) reveals that there is no significant improvement for the vector drawing tasks (DV) or items related to interpreting the longitudinal density of vectors (LD). However, in both groups, the interpretation of the transverse density of vector arrows (TD) improved significantly.

In the "unirepresentational" group, significantly better results were achieved in the items related to the interpretation of arrow lengths (AL) in the post-test compared to the pretest. In contrast, these changes were not significant in the "multirepresentational" group. The results for the individual items are given in Tables 12–17.

The tables show that when assessing the strength of the field, the participants mostly paid attention to how many vectors were drawn per spatial area. Surprisingly, this interpretation also persists in the post-test results even though this misconception was explicitly addressed during the intervention.

| | Pretest | | Post-test | | *z* | *p* | *p_korr* | *r* |
|---|---|---|---|---|---|---|---|---|
| | Mean | SD | Mean | SD | | | | |
| DV | 7.42 (82.4 %) | 2.872 (31.9 %) | 8.05 (89.4 %) | 2.398 (26.6 %) | 2.385 | <0.05 | >0.05 <0.10 | 0.30 |
| TD | 0.26 (13.0 %) | 0.599 (30.0 %) | 0.47 (23.5 %) | 0.718 (35.9 %) | 2.504 | <0.05 | <0.05* | 0.32 |
| LD | 0.69 (34.5 %) | 0.916 (45.8 %) | 0.85 (42.5 %) | 0.921 (46.0 %) | 1.618 | >0.05 | >0.05 | 0.21 |
| AL | 0.92 (46.0 %) | 0.893 (44.7 %) | 1.05 (52.5 %) | 0.931 (46.6 %) | 0.722 | >0.05 | >0.05 | 0.09 |

Table 10. Pre-post comparison of the results in generating and interpreting vector plots in Group “Multirepresentational”.

| | Pretest | | Post-test | | *z* | *p* | *p_korr* | *r* |
|---|---|---|---|---|---|---|---|---|
| | Mean | SD | Mean | SD | | | | |
| DV | 7.77 (86.3 %) | 2.378 (26.4 %) | 8.27 (91.9 %) | 1.416 (15.7 %) | 2.067 | <0.05 | >0.05 <0.10 | 0.26 |
| TD | 0.29 (14.5 %) | 0.611 (30.1 %) | 0.66 (33.0 %) | 0.848 (42.4 %) | 3.277 | <0.01 | <0.05* | 0.42 |
| LD | 0.65 (32.5 %) | 0.870 (43.5 %) | 0.87 (43.5 %) | 0.932 (46.6 %) | 1.841 | >0.05 | >0.05 | 0.23 |
| AL | 0.92 (46.0 %) | 0.893 (44.7 %) | 1.37 (68.5 %) | 0.854 (42.7 %) | 3.019 | <0.01 | <0.05* | 0.38 |

Table 11. A pre-post comparison of the results for generating and interpreting vector plots in the “unirepresentational” group.

| | TD1 | TD2 |
|---|---|---|
| Correct | 0.16 | 0.11 |
| Low Density | 0.03 | 0.03 |
| High Density | 0.77 | 0.81 |
| None | 0.02 | 0.05 |

Table 12. Results of the individual items in the pretest related to interpreting the transverse density of vector arrows. The table contains the fractions of answers that were correct (“Correct”), those that judged that the field with a lower/higher transverse density of vectors is stronger (“Low Density”/“High Density”), or those that stated that neither answer was correct (“None”).

| | LD1 | LD2 |
|---|---|---|
| Correct | 0.35 | 0.31 |
| Low Density | 0.01 | 0.02 |
| High Density | 0.60 | 0.65 |
| None | 0.02 | 0.02 |

Table 13. Results of the individual items in the pretest related to interpreting the longitudinal density of vector arrows. As above, the table contains the fractions of answers that were correct ("Correct"), those that judged that the field with a lower/higher longitudinal density of vectors is stronger ("Low Density"/"High Density"), or those that stated that neither answer was correct ("None").

| | AL1 | AL2 |
|---|---|---|
| Correct | 0.50 | 0.42 |
| Short Arrow | 0.02 | 0.05 |
| Both | 0.40 | 0.45 |
| None | 0.06 | 0.08 |

Table 14. Results of the individual items in the pretest related to interpreting the length of vector arrows. The fractions of answers that were correct ("Correct"), that considered the short arrows to be stronger ("Short Arrow"), that judged that both fields to be equally strong ("Both"), and that no answer was correct ("None") are shown.

| | Multirepresentational | | Unirepresentational | |
|---|---|---|---|---|
| | TD1 | TD2 | TD1 | TD2 |
| Pretest | | | | |
| Correct | 0.15 | 0.11 | 0.18 | 0.11 |
| Low Density | 0.06 | 0.05 | 0.00 | 0.02 |
| High Density | 0.77 | 0.79 | 0.77 | 0.82 |
| None | 0.00 | 0.05 | 0.05 | 0.05 |
| Post-test | | | | |
| Correct | 0.32 | 0.15 | 0.35 | 0.31 |
| Low Density | 0.06 | 0.05 | 0.05 | 0.03 |
| High Density | 0.60 | 0.74 | 0.53 | 0.63 |
| None | 0.02 | 0.06 | 0.03 | 0.02 |

Table 15. Results of the individual items in the pretest and post-test related to interpreting the transverse density of vector arrows in both intervention groups.

| | Multirepresentational | | Unirepresentational | |
|---|---|---|---|---|
| | LD1 | LD2 | LD1 | LD2 |
| Pretest | | | | |
| Correct | 0.37 | 0.32 | 0.34 | 0.31 |
| Low Density | 0.02 | 0.03 | 0.00 | 0.02 |
| High Density | 0.58 | 0.63 | 0.61 | 0.66 |
| None | 0.02 | 0.02 | 0.03 | 0.02 |
| Post-test | | | | |
| Correct | 0.47 | 0.39 | 0.48 | 0.39 |
| Low Density | 0.03 | 0.03 | 0.02 | 0.02 |
| High Density | 0.45 | 0.55 | 0.44 | 0.53 |
| None | 0.05 | 0.03 | 0.05 | 0.05 |

Table 16. Results of the individual items in the pretest and post-test related to interpreting the longitudinal density of vector arrows in both intervention groups.

| | Multirepresentational | | Unirepresentational | |
|---|---|---|---|---|
| | AL1 | AL2 | AL1 | AL2 |
| Pretest | | | | |
| Correct | 0.53 | 0.39 | 0.47 | 0.45 |
| Short Arrow | 0.03 | 0.06 | 0.02 | 0.03 |
| Both | 0.39 | 0.50 | 0.42 | 0.40 |
| None | 0.05 | 0.05 | 0.08 | 0.11 |
| Post-test | | | | |
| Correct | 0.58 | 0.47 | 0.73 | 0.65 |
| Short Arrow | 0.00 | 0.03 | 0.03 | 0.00 |
| Both | 0.37 | 0.47 | 0.16 | 0.29 |
| None | 0.05 | 0.03 | 0.06 | 0.06 |

Table 17. Results of the individual items in the pretest and post-test related to interpreting the length of vector arrows in both intervention groups.

5. Multirepresentational vs. unirepresentational exercises to help identify areas of maximum magnitude of field strength from unit vector representations of known field configurations

In addition to the above analyses, we tested whether learners can identify areas of maximum magnitude of the field strength from conventional unit vector representations of previously known field configurations.

The mean score in the "multirepresentational" intervention group increased from 0.74 (37.0 %), SD: 0.922 (46.1%) in the pretest to 0.79 (39.5%), SD: 0.960 (48.0%) in the post-test. Similarly, the score in the "unirepresentational" group increased from 0.90 (45.0 %), SD: 0.987 (49.4%) to 0.98 (49.0%), SD: 0.983 (49.2%). The mean difference score between the pretest and post-test was 0.0484 (2.4%), SD: 0.52565 (26.3%) in the "multirepresentational" group and 0.0806 (4.0%), SD: 0.79545 (39.8%) in the "unirepresentational" group.

No significant effect could be found for the interventions in the "multirepresentational" group ($z$=0.722) or the "unirepresentational" group ($z$=0.907). There were also no significant differences identified in the difference scores of the pre- and post-test results between the two intervention groups ($U$=1914.0, $Z$=-0.062, $p$ >0.05).

A closer examination of the answers to the individual field strength-related items reveals that the proportion of correct answers changes only slightly (see Table 19). However, the data reveal a tendency that, after the exercise, more participants state that the field is equally strong everywhere, while the proportion who interpret that the field is strongest in the location containing the most vector arrows decreases.

| Pretest | FS1 | FS2 |
|---|---|---|
| Correct | 0.44 | 0.39 |
| Less | 0.02 | 0.05 |
| More | 0.42 | 0.46 |
| Equal | 0.09 | 0.08 |

Table 18. Pretest results of the FS items. This table lists the fractions of correct answers ("Correct"), answers that estimated that the field is stronger in the area with less/more vector arrows ("Less"/ "More"), and answers that estimated that the field has the same strength in all three areas ("Equal").

| | Multirepresentational | | Unirepresentational | |
|---|---|---|---|---|
| | FS1 | FS2 | FS1 | FS2 |
| Pretest | | | | |
| Correct | 0.40 | 0.34 | 0.47 | 0.44 |
| Less | 0.03 | 0.10 | 0.00 | 0.00 |
| More | 0.40 | 0.44 | 0.44 | 0.48 |
| Equal | 0.11 | 0.10 | 0.06 | 0.06 |
| Post-test | | | | |
| Correct | 0.40 | 0.39 | 0.52 | 0.47 |
| Less | 0.02 | 0.06 | 0.02 | 0.05 |
| More | 0.35 | 0.32 | 0.27 | 0.27 |
| Equal | 0.23 | 0.23 | 0.16 | 0.16 |

Table 19. Relative pretest and post-test results of the FS items in both intervention groups.

# IV. Discussion

## 1. Findings regarding the interpretation of field geometries from conventional vs. centered unit vector representations

The results of this study show that learners recognize field geometries much better from centered unit vectors. The exploratory factor analysis reveals that when interpreting conventional unit vector representations, various sources of error arise that should be examined more closely in a future study with a larger sample size. Errors occur especially often when interpreting conventional unit vector representations of rotational fields—in particular, the border areas of the conventional unit vector representation are often misinterpreted.

However, there are field geometries that can be read from conventional unit vector representations equally well as from centered vector representations. In the present study, this was the case for the electric fields containing a single negative point charge (I), two negative point charges of the same magnitude and sign (II), and two negative point charges of different magnitude and the same sign (IV), i.e, fields that only contain electric field sinks. These cases share the commonality that the angle between the vectors and the line perpendicular to the border always lies within the range of approximately -45° to +45°; hence, the vectors always meet the border at a relatively steep angle. If mistakes were made in these three configurations, it was because the wrong geometry was chosen. The border areas, on the other hand, were interpreted correctly in most cases. It is expected that pure source fields would also produce a similar response behavior; however, this cannot be substantiated by the currently available data.

From the centered unit vector representations, the geometries of the fields and borders of the measuring area were consistently interpreted very well. If errors occurred, it was because the field geometry was not correctly assessed, whereas the borders of the plots rarely caused any interpretation issues. Although this convention for representing the directions of vector fields is not usually taught in schools, it intuitively leads to better results than the traditional approach.

## 2. Discussion of the results of the cross-representational exercise and the exercise which argues within the conventional representation to improve unit vector interpretation

Both interventions led to significantly better results for the interpretation of conventional unit vectors in the post-test. When interpreting centered unit vectors, however, only the intervention that related the conventional and centered unit vector representations to each other led to a significant improvement. Therefore, creating a relationship between the two representations not only improved the participants' interpretation of conventional unit vectors but also of the centered ones. The intervention, which argued within the representation with conventional unit vectors in its explanation did not significantly deepen understanding of the items regarding centered vectors.

Consequently, this can also be demonstrated by comparing the difference scores from the pretest and the post-test. In terms of the conventional vector representations, both training methods were similarly effective, and there were no significant differences in score improvement from the pretest to the post-test. In contrast, for the interpretation of centered unit vectors, the intervention that related both representations led to a significantly greater improvement.

These findings thus indicate that a promising approach to deepen students' understanding of conventional unit vectors and centered unit vectors would be to use an exercise that contrasts both representations and explains how to translate from one representation to the other rather than describing only the key elements of a single representation type. The effectiveness of this approach may relate to both representations being equally deepened during practice; thus, not only the conventional unit vector representation is practiced, but the centered unit vectors are also better understood. However, it should be noted that potential ceiling effects may have occurred when interpreting the centered vectors, as very good results were already achieved in the pretest.

For the conventional unit vectors, the results suggest that it is particularly difficult for learners to correctly interpret the field at the border of the vector representation. Both interventions resulted in significant improvements in this skill.

3. Insights regarding the generation and interpretation of vector representations

Almost all test subjects solved the items for generating vector representations correctly, while significant difficulties were observed when interpreting vector representations. The factor analysis showed virtually no correlation between the factor related to drawing vector representations and the factors related to interpreting vector representations. Even if the conventions of vector representations can be used correctly for drawing, it cannot be assumed that this knowledge will also be applied correctly to interpreting the vector representations. Drawing and interpreting vector representations therefore appear to be different skills that must be practiced intensively and in an integrated way.

No significant effect on the drawing of vector representations was demonstrated in either intervention. However, this could also relate to the pretest scores in this category being already very high.

Difficulties arose in the interpretation of the longitudinal density of vectors both in the pretest and post-test. No significant improvement was achieved with either intervention, and no differences were found between the two groups.

When interpreting the length of vectors, the exercise that argues within a single representation resulted in a significant interpretation improvement. This probably occurred because the learners spent more time on conventional vector notation during this exercise and less time on transfer between conventional vectors and centered vectors.

In contrast, no significant improvement was identified between the pretest and post-test in the comprehensive exercise that relates conventional vectors to centered ones. However, no significant differences in the difference score from the pretest to the post-test could be found between the intervention groups.

Notably, the main error that arose was that learners used the density of the vectors and not their length to estimate the strength of the field. The mistake was also made very frequently in the post-test. Therefore, the effects reported by Bollen et al. [32] and Elby [34] are also demonstrated here (see I.B). Additionally, there are also indications that their results should be examined in more detail, specifically in terms of distinguishing between the transverse and longitudinal density of vectors.

The results show that it is important to investigate the interpretation of transverse and longitudinal density separately as these two concepts seem to be perceived differently by learners. This is also demonstrated by the results of the factor analysis, as there is only a weak-to-moderate correlation between the two factors that test whether the longitudinal or transverse density of vectors leads to a misinterpretation of the strength of the fields.

Both interventions had a significant positive effect on the answers to the test items dealing with the interpretation of the transverse density of vectors. The difference scores between the pretest and post-test did not differ between the two groups; thus, both exercises were similarly effective for this task. This effect can possibly be explained by the fact that after the training, the concept that an increased transverse field line density implies a stronger field is no longer simply transferred from field line representations to vector representations. However, the descriptive data show that the representation is still very often interpreted such that a higher transverse density of vectors is taken to imply a higher field strength.

Overall, apart from the aforementioned differences regarding vector length interpretation, the two exercises were almost equally effective for generating and interpreting vector representations. It can also be concluded that the topic of vector representations has not yet been completely understood by the end of the 11th grade and that the participants have a considerable need for more in-depth study.

Based on the findings of this analysis, it would be reasonable to consider using additional alternative forms of representation in schools. For example, unit vector representations in conventional and centered form with color coding could be used to clearly display information about the strength of the fields without requiring overlapping vectors. For questions that only concern the magnitude of the field strength, heatmaps, contour maps, or 3D representations could also be valuable alternatives. In the sense of creating coherence, a flexible use of all these forms of representation seems to be the basis for a deeper understanding. Therefore, the learning effectiveness of these forms of representation should be systematically examined in future studies.

4. Interpretation of the results regarding the identification of areas of maximum magnitude of field strength in unit vector representations of known fields

The pretest answers show that a high proportion of participants judged the field’s strength by evaluating the number of unit vector heads per area element. This is probably due to the visual impression of these representations, combined with the interpretation that a higher element density means a higher field strength. Interestingly, the effects reported by Bollen et al. [32]

and Elby [34] (see section I.B) are also demonstrated for unit vector representations in the present study.

After training, the results revealed that there were still difficulties in understanding. In terms of performance improvement, both training exercises had no significant effect. After the intervention a higher percentage of respondents rated the representation as having the same field strength everywhere. This indicates that participants were seemingly unable to associate the field geometry with a known field configuration. Additionally, a comprehension problem in the interpretation of unit vector representations could have arisen here due to the training.

# V. Limitations

In the used tests, some concepts were tested with only a few items. To obtain more detailed insights into each concept, a closer examination would be necessary, which was not possible within the scope of this work. The present study aimed to provide a broad overview of students' understanding of different vector representations while not exceeding the time frame available for the investigation. However, the items in this study still provide valuable insights for investigating and interpreting this research theme, as detailed above.

# VI. Conclusions

The results of the present study show that vector representations are usually drawn correctly by the end of 11th grade; however, when interpreting vector representations, stubborn problems of understanding arise. This difference between generative and interpretative tasks is particularly noteworthy, and the causal relationships between these factors should be examined more closely in the future. Consequently, from a teaching perspective, it is essential that drawing and interpreting vectors should be practiced in an integrated way.

The data show that comprehension difficulties concern both the interpretation of the directions and the strengths of vector fields. To examine the aspect of interpreting directions more closely, two types of unit vector representations were compared. The results reveal that while the directions of fields can be read very intuitively from representations with centered vectors, learners have difficulties understanding conventional directional representations. To improve the learners' understanding, the tested exercise that links the two unit vector types proved to be particularly successful. However, even after the exercise, it was difficult for learners to identify already-known field configurations from directional unit representations to compare field strengths in different areas.

Further mistakes were made when interpreting vector drawings where the information describing the field strength is encoded in the vector length. The field strength is often estimated to be stronger when the spatial density of vectors is higher. The length of the vectors often is not considered. The results were partly improved by the exercises; however, the exercises in this study were short and more classroom time should ideally be spent on understanding vector representations.

A promising approach for teaching would be the following: when evolving from force vectors on a test charge to field vectors that describe the properties of specific points in space (see section I.B), attention should be paid to explaining how the fields' directions can be read from the representation. To illustrate the procedure, and for reasons of clarity and cognitive load, unit vector representations should be preferred for this task (see section I.A). A promising approach would be to start with field configurations that can already be interpreted well from a conventional vector representation. This study’s results show that examples of such fields include the electric field of a single point charge (I), the electric field of two point charges of the same magnitude and sign (II), and the electric field of two point charges of different magnitudes and the same sign (IV). In the next step, making a connection between conventional and centered vectors via a cross-representational exercise could emphasize the meaning of the representations and help foster global coherence [40]. Subsequently, students could work on field configurations that would potentially lead to incorrect assessments in the conventional notation (e.g., rotational fields) and describe the differences in the corresponding representation with centered vectors. However, further studies are needed to validate this approach.

Since the multirepresentational exercise revealed that it can be helpful to integrate different forms of representation in one exercise, it may also be helpful to introduce and combine further different representations that accentuate individual properties of the field (e.g., heatmaps and contour maps to illustrate the field’s strength in combination with unit vector representations that provide information about the field’s directions). Learning from these multirepresentational systems can fulfill the functions of learning with multiple external representations [9] and thus lead to a deeper understanding. The results described in this article provide initial insights into learning with field vector representations and should be further expanded by systematically examining other combinations of field representations in follow-up studies.